\documentclass[preprint,prx,amsmath,groupaddress,aps,subeqn,showpacs]{revtex4}
\usepackage{graphicx}
\usepackage{amssymb}
\usepackage{bm}
\usepackage{subfigure}
\usepackage{graphpap}
\usepackage{color}

\usepackage{multirow}

\begin{document}

\title{Intrinsic Transport Properties of Electrons and Holes in Monolayer Transition Metal Dichalcogenides}

\author{Zhenghe Jin}
\affiliation{Department of Electrical and Computer Engineering,
North Carolina State University, Raleigh, NC 27695-7911}

\author{Xiaodong Li}
\affiliation{Department of Electrical and Computer Engineering,
North Carolina State University, Raleigh, NC 27695-7911}

\author{Jeffrey T. Mullen}
\affiliation{Department of Electrical and Computer Engineering, North Carolina
State University, Raleigh, NC 27695-7911}

\author{Ki Wook Kim}\email{kwk@ncsu.edu}
\affiliation{Department of Electrical and Computer Engineering, North Carolina
State University, Raleigh, NC 27695-7911} 

\begin{abstract}

Intrinsic electron- and hole-phonon interactions are investigated in monolayer transition metal
dichalcogenides MX$_2$ (M=Mo,W; X=S,Se) based on a density functional theory formalism.  Due to their structural similarities, all four materials  exhibit qualitatively comparable scattering characteristics with the acoustic phonons playing a dominant role near the conduction and valence band extrema at the K point.  However, substantial differences are observed quantitatively leading to disparate results in the transport properties.
Of the considered, WS$_2$ provides the best performance for both electrons and holes with high
mobilities and saturation velocities in the full-band Monte Carlo analysis of the Boltzmann transport equation.  It is also found that monolayer MX$_2$
crystals with an exception of MoSe$_2$ generally show hole mobilities comparable to or
even larger than the value for bulk silicon at room temperature, suggesting a potential
opportunity in p-type devices.  The analysis is extended to estimate the effective
deformation potential constants for a simplified treatment as well.

\end{abstract}

\maketitle

\section{Introduction}
Beyond graphene, two-dimensional (2D) materials have shown fascinating electrical, mechanical,
optical, and spintronic properties \cite{RefWorks:1}. Particularly, transition metal dichalcogenides (TMDs) have recently become a focus of major research efforts. Unlike graphene with all of the atoms on the same plane, each monolayer of TMDs comprises one plane of transition metal atoms sandwiched between two chalcogen planes via the covalent bonding that is arranged in a trigonal prismatic network \cite{RefWorks:1,Wilson1969}. On the other hand, the inter-layer coupling is by the weak van der Waals force  that can be broken easily layer by layer as in graphene. The transition metal atoms usually consist of Mo or W, while S, Se, or Te is selected for the chalcogen.  These materials exhibit non-zero energy gaps, consistent with the diatomic nature \cite{Johari2012,Wang2012,Friend1987}, making them potential alternatives to conventional 3D semiconductors in the highly scaled applications \cite{Jariwala2014}.

A detailed investigation of charge transport, especially the intrinsic properties, is crucial for assessing the technological significance of the material as well as for fundamental physical understanding. The mechanism that determines the intrinsic transport is the carrier interaction with the lattice vibrations, i.e., phonons.  Unfortunately, most of the experimental studies  in the literature have been carried out under a broad range of conditions (such as different TMD layer thicknesses, impurity concentrations/sample qualities, and substrates) \cite{Amani2013,Radi2013,Jariwala2014,Podzorov2004,Fang2012,Liu2013,Huang2014}.
In particular, the characteristics extracted from transistor current-voltage (I-V) measurements are indirect and can be strongly affected by extrinsic factors as well as estimation
inaccuracies \cite{Fuhrer2013,Radi2013b}. As such, the reported results vary widely even for a specific material and have so far provided only an incomplete picture of carrier-phonon scattering processes. Theoretical accounts of the subject without resorting to empirical estimates have been more limited $-$ confined essentially to the conduction band electrons in MoS$_2$ \cite{Kaasbjerg2012,Li2013}.


In this work, we extend the theoretical analysis to examine intrinsic transport of both electrons and holes in four technological relevant monolayer TMDs (MX$_2$; M=Mo,W; X=S,Se) from first principles.  The density functional theory (DFT)
formalism is applied to determine the electronic band structure as well as the phonon spectra and
carrier-phonon coupling matrix elements for all phonon branches \cite{Baroni2001,Borysenko2010}.
The obtained scattering rates are subsequently used in the Boltzmann transport equation to evaluate the intrinsic velocity-field characteristics with a full-band Monte Carlo treatment. Comparison between the considered TMDs clearly elucidates the key factors affecting carrier transport in this material system along with the potential applications.  Finally, the effective deformation potential constants are extracted for the relevant intrinsic carrier-phonon interaction processes.

\section{Theoretical Model}
In the scattering calculations, each phonon is treated as a perturbation of the self-consistent potential within the linear response; in other words, via density functional perturbation theory
(DFPT) \cite{Baroni2001}. Then the elements of the electron-phonon interaction matrix can be written as:
\begin{equation}
g_{\textbf{q},\textbf{k}}^{(i,j)\nu}=\sqrt{\frac{\hbar}{2M\omega_{\nu,\textbf{q}} }}  \langle j,
\textbf{k}+\textbf{q}|\Delta V_{\textbf{q},\mathrm{SCF}}^\nu|i,\textbf{k} \rangle ,
\end{equation}
where $|i,\textbf{k} \rangle $ is the Bloch eigenstate with wave vector $\textbf{k}$, band index $i$, and energy $E_{ \mathbf{k}}^i$; $\Delta V_{\textbf{q},\mathrm{SCF}}^\nu$ is the derivative of the self-consistent Kohn-Sham potential with respect to the atomic displacement of a phonon from branch $\nu$ with wave vector $\mathbf{q}$ and frequency $\omega_{\nu, \mathbf{q}}$; and $M $ is the atomic mass.
As mentioned earlier, the scattering matrix elements can be evaluated from the DFT/DFPT framework along with the electronic and phononic dispersion relations.  The actual computation is performed with the QUANTUM ESPRESSO package using the norm conserving pseudopotentials and the local density approximation for the exchange-correlation functional \cite{Giannozzi2009}. For the k-point sampling, a 18$\times$18$\times$1 Monkhorst-Pack grid is used with the energy cut-off of 60 Ry in the plane-wave expansion.

With the matrix elements obtained in the first Brillouin zone (FBZ), the scattering rates can be calculated by using the Fermi's golden rule:
\begin{equation}
\frac{1}{\tau_\textbf{k}^i}=\frac{2\pi}{\hbar} \sum_{\textbf{q},\nu, \pm}
|g_{\textbf{q},\textbf{k}}^{i,\nu}|^2
(N_{\nu,\textbf{q}}+\frac{1}{2} \pm \frac{1}{2})\delta (E_{\textbf{k}\mp \textbf{q}}^i
\pm \hbar \omega_{\nu,\textbf{q}}-E_{\textbf{k}}^i) ,
\end{equation}
where $\pm$ represents phonon emission and absorption, respectively, and $N_{\nu,\textbf{q}}$ is the phonon occupation number following the Bose-Einstein statistics.  Index $j$ is dropped hereinafter as the interband transitions are not considered. Further, the present analysis concerns only the lowest conduction
band ($i=c$) and the highest valence band ($i=v$) for the electrons and holes, respectively, with the assumption of non-degeneracy (i.e., no dependence on the Fermi level).  In the case of holes, a slight modification is needed in Eq.~(2) since the scattering matrix elements are formulated from the view point of electrons. In other words, the computed matrix element for the $\textbf{k}$ to $\textbf{k}^\prime$ transition actually corresponds to the $-\textbf{k}^\prime$ to $-\textbf{k}$ scattering for holes in the valence band; note that the hole state momentum is, by convention, defined with the opposite polarity to that of the corresponding electronic state.  The DFT/DFPT results provide the physical details necessary to solve the Boltzmann transport equation for transport characteristics with the Monte Carlo technique \cite{Li2013,Borysenko2010}.


\section{Results and Discussion}

\subsection{Electronic and phononic dispersion}

The DFT calculation of monolayer MX$_2$ shows direct band gaps with the conduction and valence band extrema at the K point in the momentum space for all combinations of M and X under consideration. This result is consistent
with other theoretical calculations as well as experimental measurements (see, for example, Ref.~\onlinecite{Chang2013}). For the lowest conduction band, a satellite energy minimum is also observed along the K-$\Gamma$ axis at nearly the half-way point that is often called the Q valley [see Fig.~1(a)]. On the other hand, the secondary peak for the valence band is located at the Brillouin zone center
(i.e., the $\Gamma$ point).  Table~I summaries the optimized lattice constants, Q-K and $\Gamma$-K separation energies
for the conduction ($ E_\mathrm{Q K}^c $) and valence bands ($ E_\mathrm{\Gamma K}^v $), respectively, as well as the corresponding effective masses.  Notably, the estimates for $ E_\mathrm{Q K}^c $ ($\lesssim 80$ meV) are in the range that could significantly affects electron transport via the intervalley scattering even under low bias
conditions, whereas those of $ E_\mathrm{\Gamma K}^v $ are generally sizable ($\gtrsim 150$ meV).
Comparison with similar calculations in the literature \cite{Kaasbjerg2012,Chang2013} suggests that the intervalley separation energies are sensitive to the details of the first principles approach including the pseudopotentials.
In particular, application of the local density approximation tends to predict smaller values, requiring further clarification via, perhaps, accurate experimental determination.  As for the effective masses, both the longitudinal and transverse components are needed for the Q valleys due to the anisotropic nature. Those listed on the table represent the geometric mean (corresponding to the density-of-states effective mass except the degeneracy factor), which happens to be very close to the conductivity effective masses as well.

With three atoms in a unit cell, monolayer MX$_2$ has the symmetry of the point group $D_{3h}$ and nine phonon branches. The irreducible representations of each phonon mode, along with the polarization, identify the different vibrational dynamics as in Fig.~1(b), where WS$_2$ is considered a prototypical example.  For instance, the lowest three are the longitudinal acoustic (LA), transverse acoustic (TA), and out-of-plane acoustic (ZA) modes.
Both the LA and TA modes show linear dispersion in the long wavelength limit, while the ZA phonons deviate with an approximate quadratic dependence.  The estimated longitudinal sound velocity (i.e., slope) is 4.3$\times$10$^5$ cm/s
for monolayer WS$_2$. The next two are the in-plane optical E$^{\prime\prime}$ modes, with two S atoms vibrating out of  phase and the W  atomic static.  Then, the E$^\prime$ branches are polar LO and TO modes, with all three atoms moving out of phase in the plane.  The polar LO-TO splitting
at the $\Gamma$ point is too small to be visible.  The highest A$_{2}^{\prime\prime}$ and A$_1$ phonons are out-of-plane optical vibrations.  More specifically, the A$_1$ mode has the W atom static while the S atoms moving in the opposite directions. All three atoms oscillate out of the plane and out of phase in the case of A$_{2}^{\prime\prime}$ phonons.
The phonon energies at different symmetry points are summarized in Table~II for MoSe$_2$, WS$_2$, and WSe$_2$; the corresponding information for MoS$_2$ can be found in Ref.~\onlinecite{Li2013}.

\subsection{Scattering rates and transport properties}

While the carrier-phonon interactions in the present study are limited to the intraband events within the lowest conduction and the highest valence bands, respectively, a number of transition possibilities exist due to the multi-valley nature of the band structure.  Figure~2 schematically illustrate the available final states via the intervalley scattering for the conduction electrons and the valence holes.  Electrons in the K minimum can scatter to the nearby K$^\prime$, Q, and Q$^\prime$ valleys, the latter two with three-fold degeneracy [Fig.~2(a)].  Similarly, those from a Q minimum can transfer to other Q  and K/K$^{\prime}$ valleys but with lower symmetry [Fig.~2(b)].  In the case of holes, the picture is simpler as the satellite peak (or the valley from the hole point of view) at $\Gamma$ does not have any equivalent point in the FBZ [Figs.~2(c,d)].

The carrier-phonon scattering rates are calculated as a function of electron/hole energy following the Fermi's golden rule.  Again using monolayer WS$_2$ as an example, Figs.~3(a,b) illustrate the characteristic phonon emission and absorption rates for electrons in the K valleys at room temperature, while the corresponding results for holes are given in Figs.~3(c,d).  For both types of carriers, the wave vector $\mathbf{k}$ of the initial state  is chosen along the K-$\Gamma$ axis.  The pronounced abrupt jumps in the scattering rates indicate the onset of intervalley transitions, which appears at higher energies in the valence band consistent with the obtained band structure (i.e., $ E_\mathrm{\Gamma K}^v >  E_\mathrm{Q K}^c $). It is also clear from the figure that the acoustic phonons provide the dominant scattering mechanism for electrons as well as holes.  One difference is that the LA mode is the largest in the conduction band, whereas it is the TA vibration for the valence band. The role of optical phonons is relatively minor particularly at low carrier energies.  Of nine phonon branches, only LA, TA, E$^\prime$ (LO and TO), and A$_1$ modes are relevant to the scattering with electrons or holes.  The influence of the remaining four branches is negligible due to the weak coupling \cite{Li2013}.  The qualitative features of mode specific carrier-phonon interactions are essentially similar in all four TMDs under consideration.

For a more quantitative comparison between the TMDs, the contributions of the five branches are summed and the resulting total emission and absorption rates are plotted in Fig.~4.  Two points are apparent from the analysis.  In the choice of transition metal elements, the Mo compounds  (MoX$_2$) may be electrically more resistive than those of W (WX$_2$) with generally higher scattering rates for both electrons and holes.  Between the two chalcogens (i.e., MS$_2$ vs.\ MSe$_2$), it is Se that appears to scatter the carriers more strongly via the interaction with the phonons.  The latter observation concerning Se may be attributed in part to the small Q-K separation $E_\mathrm{Q K}^c $ in the conduction band compared to that of S cases that is evident from to the low onset energies of intervalley scattering.  At higher energies where phonon emission dominates, the differences in the electron scattering rates are much less pronounced.  In contrast, the hole scattering rates in MSe$_2$ is consistently higher even with the larger $\Gamma$-K separation $ E_\mathrm{\Gamma K}^v$.  Overall, MoSe$_2$ shows the highest scattering rates for both electrons and holes, while WS$_2$ provides the opposite.

Utilizing the scattering rates, the velocity versus field relation is obtained by a full-band Monte Carlo simulation at different temperatures.  Figure~5 presents the results for WS$_2$.  Needless to say, the low-field mobility and the saturation velocity show strong temperature dependence that is  characteristic of a phonon-limited process.  One potentially interesting observation is that valence hole transport could be comparable, if not superior, to the properties of conduction electrons.  For instance, the estimated intrinsic mobility is approximately 320 cm$^2$/Vs and 540 cm$^2$/Vs for electrons ($\mu_c$) and holes ($\mu_v$), respectively; the saturation velocity at 100 kV/cm shows a similar trend with $ 3.7 \times 10^6$ cm/s (electrons) and $ 4.1 \times 10^6$ cm/s (holes).  Since the conduction and valence band effective masses in WS$_2$ are fairly alike at the K point (actually, $m_\mathrm{K}^c < m_\mathrm{K}^v$ according to Table~I), the advantage of hole transport can be understood by the low scattering rates.  As can be seen from Fig.~3, holes experience generally less scattering with phonons throughout the presented energy range.  One point to bear in mind is the possibility that our DFT band calculation may have underestimated $E_\mathrm{Q K}^c $.  When the impact of K-Q intervalley scattering is removed (corresponding to a large separation energy), the intrinsic mobility for the K-valley electrons ($\mu_{c,\mathrm{K}}$) increases to about 690 cm$^2$/Vs.  Nonetheless, it is fair to say that hole transport shows a comparative advantage in WS$_2$ as electrons have traditionally been the faster carrier in conventional semiconductors.

Table~III summarizes the calculated transport properties of the considered TMDs.  Note that two values are provided for the electron saturation velocity $v_{sat}^c$.  The smaller number provides the drift velocity obtained at 100 kV/cm, while the other represents an extrapolation to higher fields for the saturation behavior.  As indicated in Fig.~5, the velocity-field curves for the conduction band electrons show an upward slope even at 100 kV/cm.  In comparison, the hole drift velocities reach the plateau $v_{sat}^v$ earlier at approximately 50$\sim$60 kV/cm.
Comparison between the TMDs indicates the trend that is consistent with the observation on the scattering rates.  Namely, WS$_2$ and MoSe$_2$ provide the most and the least promising performances, respectively, while MoS$_2$ and WSe$_2$ are in between with comparable characteristics except $\mu_c$.  WS$_2$, in particular, also takes advantage of lighter effective masses $m_\mathrm{K}^c$, $m_\mathrm{K}^v$.  In fact, the conventional criterion  based on the ratio of phonon energy over the effective mass (i.e., $\omega / m_\mathrm{K}^{c,v}$) appears to hold in explaining the qualitative ordering of saturation velocities;  $ \mathrm{WS_2} > \mathrm{MoS_2} > \mathrm{WSe_2} > \mathrm{MoSe_2}$.
All of the materials (with a possible exception of MoSe$_2$) offer competitive transport performances of holes and could be promising alternatives to the conventional semiconductors in the p-channel applications.  By contrast, the intrinsic hole mobility of bulk Si is around 450 cm$^2$/Vs at room temperature.

With the recent flurry of activities focused on the fabrication of TMD based field effect transistors, it is interesting to compare our estimates of intrinsic mobility with the experimentally deduced values even though the channel mobility extracted from the I-V curves is not a direct measurement as discussed above.  Thus far, the most studied materials appear to be n-type MoS$_2$ and p-/n-type WSe$_2$.  As for n-type monolayer MoS$_2$, the electron channel mobility as high as approx.\ 60 cm$^2$/Vs at room temperature has been reported in the literature \cite{Huang2014}.  This number is consistent with our prediction of the intrinsic upper bound in Table~III (i.e., $\mu_c = 130$ cm$^2$/Vs, $\mu_{c, \mathrm{K}} = 320 $ cm$^2$/Vs).  Similarly, the case of p-type monolayer WSe$_2$ shows the best hole mobility around 250 cm$^2$/Vs that is comparable to the theoretical estimate of 270 cm$^2$/Vs at 300 K.  On the other hand, the values over 100 cm$^2$/Vs has been observed in the WSe$_2$ n-channels \cite{Liu2013}, which is actually higher than $\mu_c = 30$ cm$^2$/Vs (but still lower than $\mu_{c, \mathrm{K}} = 250 $ cm$^2$/Vs).  This anomaly (i.e., an experimental mobility larger than $\mu_c$) strongly suggests that our DFT calculation may have indeed underestimated the intervalley separation $E_\mathrm{Q K}^c $, leading to substantial overestimation of intervalley scattering via the low onset energy. Once the overly strong K-Q electron transfer is discounted, the resulting K-valley dominated mobility $\mu_{c, \mathrm{K}}$ is certainly more consistent with the experimentally extracted.  Considering the uncertainties in accurate evaluation of $E_\mathrm{Q K}^c $ in TMDs mentioned previously, $\mu_{c, \mathrm{K}}$ appears to provide a more reliable estimate than $\mu_c$ for the theoretical upper bound of the electron mobility.  When comparison is made with the cases employing non-monolayers, extra caution is required as both the conduction band and valence band valleys/peaks undergo changes in energy (e.g., the minimum/maximum points shifting in the momentum space).


An additional note before moving to the next discussion is that our DFT/DFPT results described thus far have been obtained with a non-relativistic pseudopotential that does not consider the effect of spin-orbit coupling.  However, it is known that this interaction can be strong in TMDs with a sizable band splitting, particularly in the valence band, through the lifted spin degeneracy \cite{Xiao2012}.  For instance, WS$_2$ may experience valence band splitting of approximately 450 meV at the K point while the corresponding number for the conduction band is only about 30 meV.  We examine the impact of spin-orbit coupling on the carrier-phonon scattering by adopting both the non-relativistic and relativistic pseudopotential.  The DFPT calculation shows that the scattering matrix elements remain largely unchanged as illustrated in Fig.~6;  the effect of spin split bands is essentially through the modified density of final states that can be accommodated by properly conducting the sum/integral in Eq.~(2).  With the consideration that the spin flipping events are not the primary scattering process near the band extrema, use of a non-relativistic pseudopotential appears adequate in the current study of electrical properties.

\subsection{Deformation potential approximation}

For practical applications, it would be convenient if the \emph{first principles} results
can be approximated by a simple analytical model.  Particularly useful in the carrier-phonon interaction is the deformation potential approximation.  Under this treatment, the scattering potential
$\Delta V_{\textbf{q},\mathrm{SCF}}^\nu$ can be simplified by the expressions in the zeroth order ($D_0$) or the first order (D$_1$q) of phonon momentum q \cite{Ferry2000}.  The first-order deformation potential constant ($D_1$) is adopted to represent the coupling matrices for the acoustic phonon modes in the long wavelength limit (i.e., intravalley scattering). In comparison, those involving the near zone-edge acoustic phonons (i.e., intervalley scattering) are treated by using the zeroth-order deformation potential ($D_0$) in a manner analogous to the optical modes. In the latter case ($D_0$), the phonon energy is assumed independent of the momentum for simplicity. The obtained analytical expressions of the scattering rates (that also utilize the effective masses) are then matched to the first-principles results by fitting the effective deformation potential constants \cite{Li2013}.

In the case of intravalley acoustic phonon scattering, the final expression for the combined TA/LA contribution gives
\begin{equation}
\frac{1}{\tau_{\textbf{k}}^{i,n}}=\frac{m_{n}^i D_1^2 k_BT}{\hbar^3\rho v_s^2} ,
\end{equation}
where $i$ denotes the band index as before ($c$,$v$), $n$  the specific valley/peak (e.g., K, Q, $\Gamma$), $m_{n}^i$ the density-of-states effective mass in the corresponding valley/peak, $\rho$ the 2D mass density, and  $v_s$ the sound velocity (for which the estimate for the LA phonon $v_\mathrm{LA}$  is used for convenience). The deformation potential constant $D_1$ must also have indices $i,n$ while not shown explicitly. On the other hand,
the optical or intervalley acoustic phonon scattering rates reduce to
\begin{equation}
\frac{1}{\tau_{\textbf{k},\nu}^{i,m,n}}=\frac{g_{n}^i m_{n}^i D_0^2}{2 \hbar^2\rho \omega_\nu}[N_\nu\Delta_1+(N_\nu+1)\Delta_2] .
\end{equation}
Here, indices $i$,$m$,$n$ specify the conduction or valence band, the initial valley/peak, and the final valley/peak, respectively, $g_n^i$ is the ($i$,$n$)-valley degeneracy, and $N_\nu$ denotes the Bose-Einstein distribution of phonon mode $\nu$. In addition, $\Delta_1$ and $\Delta_2$ formally represent the appropriate Heaviside step functions for absorption and emission processes whose onset energies account for the associated intervalley separation.  The effect of weak Fr\"{o}hlich scattering by the LO(E$^\prime$) mode is implicitly added to the deformation potential interaction [i.e., Eq.~(4)] even though the two mechanisms are of different origin.

The extracted deformation potential values are listed in Tables~IV and V for the electron and hole scattering, classified also by the initial/final states and the involved phonon momentum (including its location in the FBZ). For simplicity, the actual value  used in the analytical calculation for $ \omega_\nu$ [see Eq.~(4)] is approximated by an average of the relevant phonon modes. Specifically, the acoustic phonon energy is obtained as the mean value of LA and TA modes at the respective symmetry points, whereas the optical phonon frequency is constructed from TO(E$^\prime$), LO(E$^\prime$), and A$_1$ branches (see Table~II). Further simplification of the model can be attempted by combining multiple contributions into one effective process.  The deduced deformation potential constants allow straightforward recalculation of the scattering rates as well when the band parameters (e.g., $E_\mathrm{Q K}^c $ or $E_\mathrm{\Gamma K}^v $) require adjustments.

\section{Summary}
A first principles calculation of carrier-phonon interaction combined with the full-band Boltzmann equation is used to analyze the intrinsic transport properties of electrons and holes in TMDs.  Of the four materials under investigation, WS$_2$ and MoSe$_2$ appear to provide the most and the least promising performances, respectively, while MoS$_2$ and WSe$_2$ are in between with comparable characteristics.  More specifically, the W compounds may be electrically less resistive than those of Mo with generally lower scattering rates for both electrons and holes.  Between the two chalcogens, it is S that appears to scatter the carriers less strongly via the interaction with the phonons.  All of the materials (with a possible exception of MoSe$_2$) are predicted with competitive hole transport properties compared to bulk silicon (e.g., 540 cm$^2$/Vs in WS$_2$ vs.\ 450 cm$^2$/Vs in Si), offering attractive alternatives to conventional semiconductors in the electronic applications with ultrasmall dimensions.

\begin{acknowledgments}
This work was supported, in part, by SRC/NRI SWAN.
\end{acknowledgments}

\clearpage
\begin{table}[ht]
\setlength{\tabcolsep}{12pt}
\caption{Lattice constant, Q-K energy separation, and $\Gamma$-K energy separation of TMDs along with the effective masses at the relevant energy valleys/peaks.   Symbols $c$ and $v$ denote the conduction and valence bands, respectively. The effective masses are given in units of electron rest mass $m_0$. } 
\vspace{1cm}
\centering 
\begin{tabular}{c c c c c c c c} 
\hline\hline 
&  Lattice & $E_\mathrm{Q K}^c $ & $E_\mathrm{\Gamma K}^v $ & $m_\mathrm{K}^c$ & $m_\mathrm{Q}^c$ & $m_\mathrm{K}^v$ & $ m_\Gamma^v$ \\ [-6pt]
& const ($\text{\AA}$) & (meV) & (meV) & ($m_0$) & ($m_0$) & ($m_0$) & ($m_0$)  \\
\hline 
MoS$_2$  & 3.14 & 81 & 148 & 0.51 & 0.76 & 0.58 & 4.05  \\ 
MoSe$_2$ & 3.27 & 28 & 374 & 0.64 & 0.80 & 0.71 & 7.76  \\
WS$_2$ & 3.10 & 67 &  173 & 0.31 & 0.60 & 0.42 & 4.07 \\
WSe$_2$& 3.25 & 16 & 427 & 0.39 & 0.64 & 0.51 & 7.77  \\ [1ex] 
\hline 
\end{tabular}
\end{table}

\clearpage

\begin{table}[ht]
\setlength{\tabcolsep}{8pt}
\caption{Frequencies (in units of meV) of the phonon modes relevant to the carrier-phonon interaction at the high symmetry points in the FBZ for monolayer MoSe$_2$, WS$_2$ and WSe$_2$.  The corresponding information for MoS$_2$ can be found in Ref.~\onlinecite{Li2013}. In addition, the estimated sound velocities ($v_\mathrm{LA}$, in particular) are $6.6 \times 10^5$ cm/s, $ 4.1 \times 10^5$ cm/s, $4.3 \times 10^5$ cm/s, and $ 3.3 \times 10^5$ cm/s for MoS$_2$, MoSe$_2$, WS$_2$, and WSe$_2$, respectively.} 
\vspace{1cm}
\centering 
\begin{tabular}{  c c c c c | c c c c | c c c c}
\hline\hline 
  \multirow{2}{*}{Mode}&
 \multicolumn{4}{c }{MoSe$_2$} &
 \multicolumn{4}{c }{WS$_2$} &
 \multicolumn{4}{c }{WSe$_2$} \\
 \cline{2-13}
   & $\Gamma$ & K & M & Q & $\Gamma$ & K & M & Q &$\Gamma$ & K & M & Q   \\
 \hline
TA & 0 & 16.6 & 16.4 & 13.3 & 0 & 17.4 & 16.5 & 15.9 & 0 & 15.6 & 15.3 & 11.6 \\
LA & 0 & 19.9 & 19.7 & 16.9 & 0 & 23.6 & 22.7 & 19.5 & 0 & 18.0 & 16.3 & 14.3 \\
TO(E$^\prime$) & 36.1& 35.5 & 35.8 & 36.4 & 44.4 & 43.8 & 45.3 & 45.3 & 30.5 & 26.7 & 28.4 & 27.3 \\
LO(E$^\prime$) & 36.6 & 37.4 & 37.9 & 37.5 & 44.2 & 43.2 & 42.3 & 42.3 & 30.8 & 31.5 & 31.8 & 32.5 \\
A$_1$ & 30.3 & 25.6 & 27.3 & 27.1 & 51.8 & 48.0 & 49.8 & 50.0 &  30.8 & 31.0 & 29.8 & 30.4 \\
 \hline
 \end{tabular}
\end{table}

\clearpage
\begin{table}[ht]
\setlength{\tabcolsep}{12pt}
\caption{Calculated intrinsic mobilities and saturation velocities of TMDs at room temperature. Symbols $c$ and $v$ denote the conduction and valence bands, respectively, while $\mu_{c,{\mathrm K}}$ provides the K-valley dominated electron mobility.  For $v_{sat}^c$, two values are given; one evaluated at 100 kV/cm (smaller) and the other extrapolated to higher fields for velocity saturation (larger).  In the case of $v_{sat}^v$, the drift velocity  reaches the plateau earlier at approximately 50$\sim$60 kV/cm.} 
\vspace{1cm}
\centering 
\begin{tabular}{c c c c c c} 
\hline\hline 
& $\mu_c$ & $\mu_{c,{\mathrm K}}$ & $\mu_v$ & $v_{sat}^c$ & $v_{sat}^v$\\ [-6pt]  
& (cm$^2$/Vs) & (cm$^2$/Vs) & (cm$^2$/Vs) & (10$^6$ cm/s) & (10$^6$ cm/s) \\ 
\hline 
MoS$_2$& ~~130$^a$ & ~~320$^a$ & 270 & 3.4$^a$/4.8 & 3.8  \\ 
MoSe$_2$& 25 & 180 & 90 & 1.9/3.6 & 3.5 \\
WS$_2$& 320 & 690 & 540 & 3.7/5.1 & 4.1  \\ 
WSe$_2$ & 30 & 250 & 270 & 2.2/4.0 & 3.5 \\
[1ex] 
\hline 
$^a$Ref.~\onlinecite{Li2013}
\end{tabular}
\end{table}

\newcommand{\tabincell}[2]{\begin{tabular}{@{}#1@{}}#2\end{tabular}}
\clearpage
\begin{table}[ht]
\setlength{\tabcolsep}{12pt}
\caption{Estimated deformation potential constants for electron-phonon interaction in the lowest conduction band of monolayer MX$_2$. Symbols ac and op denote the acoustic and optical phonons, respectively.  The values given in units of eV are the first-order deformation potential $D_1$, while those in eV/cm correspond to the zeroth-order $D_0$.  See also Fig.~2 for a detailed notation of the symmetry points in the FBZ.} 
\vspace{1cm}
\centering 
\begin{tabular}{c c c c c c c} 
\hline\hline 
\tabincell{c}{Electron\\[-6pt]transition} & \tabincell{c}{Phonon mode \\[-6pt]\& momentum} & ~MoS$_2$$^a$ & MoSe$_2$ & WS$_2$ & WSe$_2$ & Units\\ [1ex] 
\hline 
\multirow{2}{*}{K $\to$ K}  & ac, $\Gamma$ & 4.5 & 3.4 & 3.2 & 3.2 & eV\\
& op, $\Gamma$ & 5.8 & 5.2 & 3.1 & 2.3 & $10^8$ eV/cm\\

\multirow{2}{*}{K $\to$ K$^\prime$}  & ac, K$^\prime$ & 1.4 & 1.8 & 1.2 & 1.3 & $10^8$ eV/cm\\
& op, K$^\prime$ & 2.0 & 2.1 & 1.1 & 0.8 & $10^8$ eV/cm\\

\multirow{2}{*}{K $\to$ Q}  & ac, Q$^\prime$ & 9.3 & 9.1 & 7.3 & 8.2 & $10^8$ eV/cm\\
& op, Q$^\prime$ & 1.9 & 1.7 & 0.9 & 0.8 & $10^8$ eV/cm\\

\multirow{2}{*}{K $\to$ Q$^\prime$}  & ac, M & 4.4 & 4.5 & 3.4 & 5.7 & $10^8$ eV/cm\\
& op, M & 5.6 & 5.3 & 2.7 & 3.2 & $10^8$ eV/cm\\
\hline 
\multirow{2}{*}{Q$_1$ $\to$ Q$_1$}  & ac, $\Gamma$ & 2.8 & 3.1 & 1.8 & 1.9 & eV\\
& op, $\Gamma$ & 7.1 & 7.8 & 3.4 & 2.7 & $10^8$ eV/cm\\

\multirow{2}{*}{Q$_1$ $\to$ Q$_2$(Q$_6$)} & ac, Q$_3$(Q$_5$) & 2.1 & 2.2 & 1.7 & 2.7 & $10^8$ eV/cm\\
& op, Q$_3$(Q$_5$) & 4.8 & 4.3 & 2.3 & 1.9 & $10^8$ eV/cm\\

\multirow{2}{*}{Q$_1$ $\to$ Q$_3$(Q$_5$)}  & ac, M$_3$(M$_4$) & 2.0 & 2.2 & 1.5 & 1.8 & $10^8$ eV/cm\\
& op, M$_3$(M$_4$) & 4.0 & 5.9 & 1.9 & 1.6 & $10^8$ eV/cm\\

\multirow{2}{*}{Q$_1$ $\to$ Q$_4$}  & ac, K$^\prime$ & 4.8 & 4.1 & 3.7 & 4.2 & $10^8$ eV/cm\\
& op, K$^\prime$ & 6.5 & 4.7 & 3.1 & 2.5 & $10^8$ eV/cm\\

\multirow{2}{*}{Q$_1$ $\to$ K}  & ac, Q$_1$ & 1.5 & 1.5 & 1.4 & 1.6 & $10^8$ eV/cm\\
& op, Q$_1$ & 2.4 & 3.0 & 1.3 & 1.0 & $10^8$ eV/cm\\

\multirow{2}{*}{Q$_1$ $\to$ K$^\prime$}  & ac, M$_2$(M$_5$) & 4.4 & 4.9 & 4.0 & 4.1 & $10^8$ eV/cm\\
& op, M$_2$(M$_5$) & 6.6 & 8.3 & 4.6 & 2.8 & $10^8$ eV/cm\\
\hline 
$^a$Ref.~\onlinecite{Li2013}
\end{tabular}
\end{table}

\clearpage
\begin{table}[ht]
\setlength{\tabcolsep}{12pt}
\caption{Estimated deformation potential constants for hole-phonon interaction in the highest valence band of monolayer MX$_2$.  The notations are the same as in Table~IV. } 
\vspace{1cm}
\centering 
\begin{tabular}{c c c c c c c} 
\hline\hline 
\tabincell{c}{Electron\\[-6pt]transition} & \tabincell{c}{Phonon mode \\[-6pt]\& momentum} & MoS$_2$ & MoSe$_2$ & WS$_2$ & WSe$_2$ & units\\ [1ex] 
\hline 
\multirow{2}{*}{K $\to$ K}  & ac, $\Gamma$ & 2.5 & 2.8 & 1.7 & 2.1 & eV\\
& op, $\Gamma$ & 4.6 & 4.9 & 2.3 & 3.1 & $10^8$ eV/cm\\

\multirow{2}{*}{K $\to$ K$^\prime$}  & ac, K$^\prime$ & 1.2 & 0.9 & 0.8 & 1.1 & $10^8$ eV/cm\\
& op, K$^\prime$ & 3.1 & 2.6 & 1.4 & 2.0 & $10^8$ eV/cm\\

\multirow{2}{*}{K $\to$ $\Gamma$}  & ac, K$^\prime$ & 1.2 & 1.5 & 1.4 & 1.7 & $10^8$ eV/cm\\
& op, K$^\prime$ & 1.9 & 1.5 & 1.9 & 1.8 & $10^8$ eV/cm\\
\hline 
\multirow{2}{*}{$\Gamma$ $\to$ $\Gamma$}  & ac, $\Gamma$ & 2.7 & 1.5 & 1.9 & 1.8 & eV\\
& op, $\Gamma$ & 3.5 & 3.8 & 1.5 & 2.2 & $10^8$ eV/cm\\

\multirow{2}{*}{$\Gamma$ $\to$ K}  & ac, K & 4.2 & 3.2 & 3.3 & 4.4 & $10^8$ eV/cm\\
& op, K & 6.1 & 5.1 & 2.9 & 5.1 & $10^8$ eV/cm\\
\hline 
\end{tabular}
\end{table}

\clearpage
\begin{center}
\begin{figure}
\includegraphics[width=8.5cm]{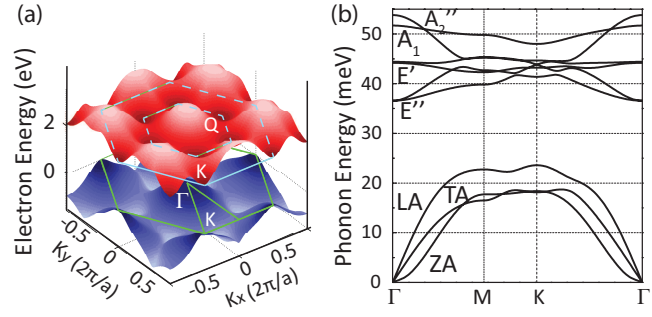} 
\caption{(Color online) (a) Electronic and (b) phononic dispersion of monolayer WS$_2$ in the FBZ.
In (a), the conduction band minimum at the K point serves as the reference of energy scale (i.e., the point of zero energy).}
\end{figure}
\end{center}

\clearpage
\begin{center}
\begin{figure}
\includegraphics[width=8cm]{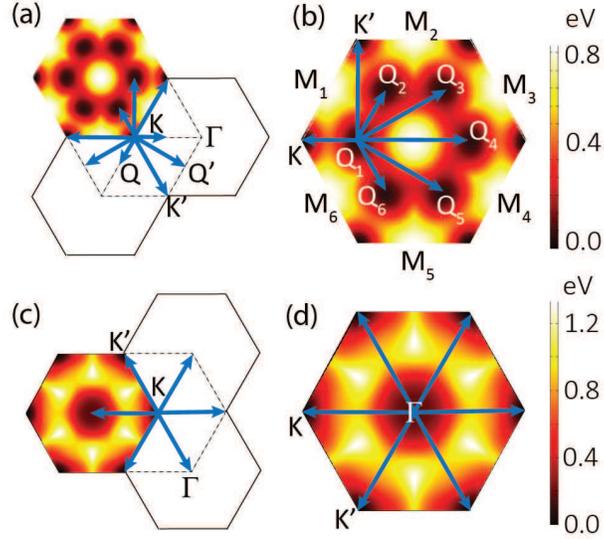} 
\caption{(Color online) Schematic illustration of intervalley scattering processes for (a) electrons in the K valley, (b) electrons in the Q valley, (c) holes in the K valley/peak, and (d) holes in the $\Gamma$ valley/peak.  The vertical color bar denotes the carrier energy with respect to the reference; i.e., zero at the bottom (top) of the conduction (valence) band for electrons (holes), respectively, at the K point.}
\end{figure}
\end{center}

\clearpage
\begin{center}
\begin{figure}
\includegraphics[width=8cm]{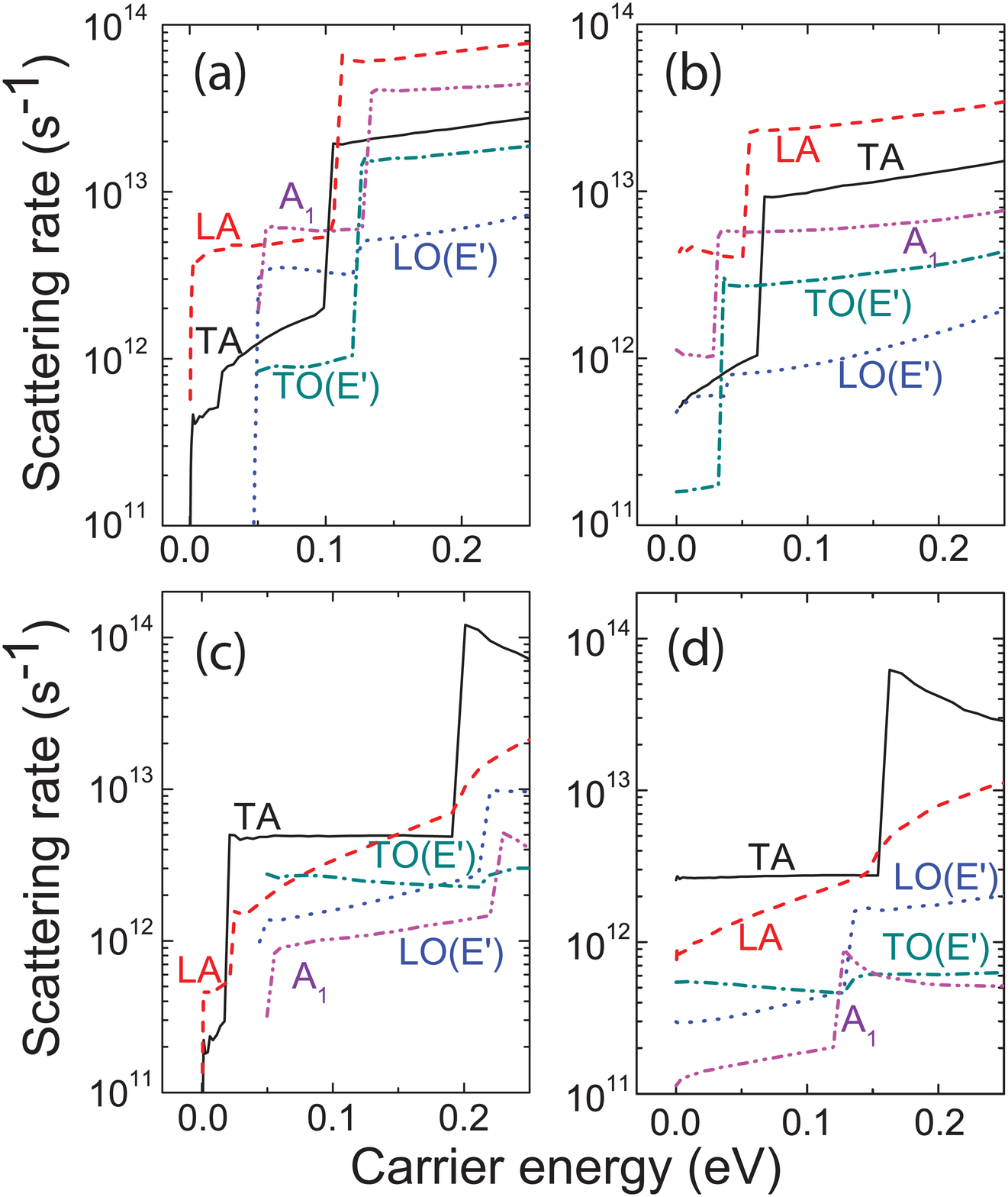} 
\caption{(Color online) Intrinsic scattering rates of (a,b) K-valley electrons and (c,d) K-valley/peak holes in monolayer WS$_2$ via different phonon modes at room temperature. (a) and (c) illustrate phonon emission, while (b) and (d) are for phonon absorption.  The electron/hole wave vector $\mathbf{k}$ is assumed to be along the K-$\Gamma$ axis as the carrier energy increases. }
\end{figure}
\end{center}

\clearpage
\begin{center}
\begin{figure}
\includegraphics[width=8cm]{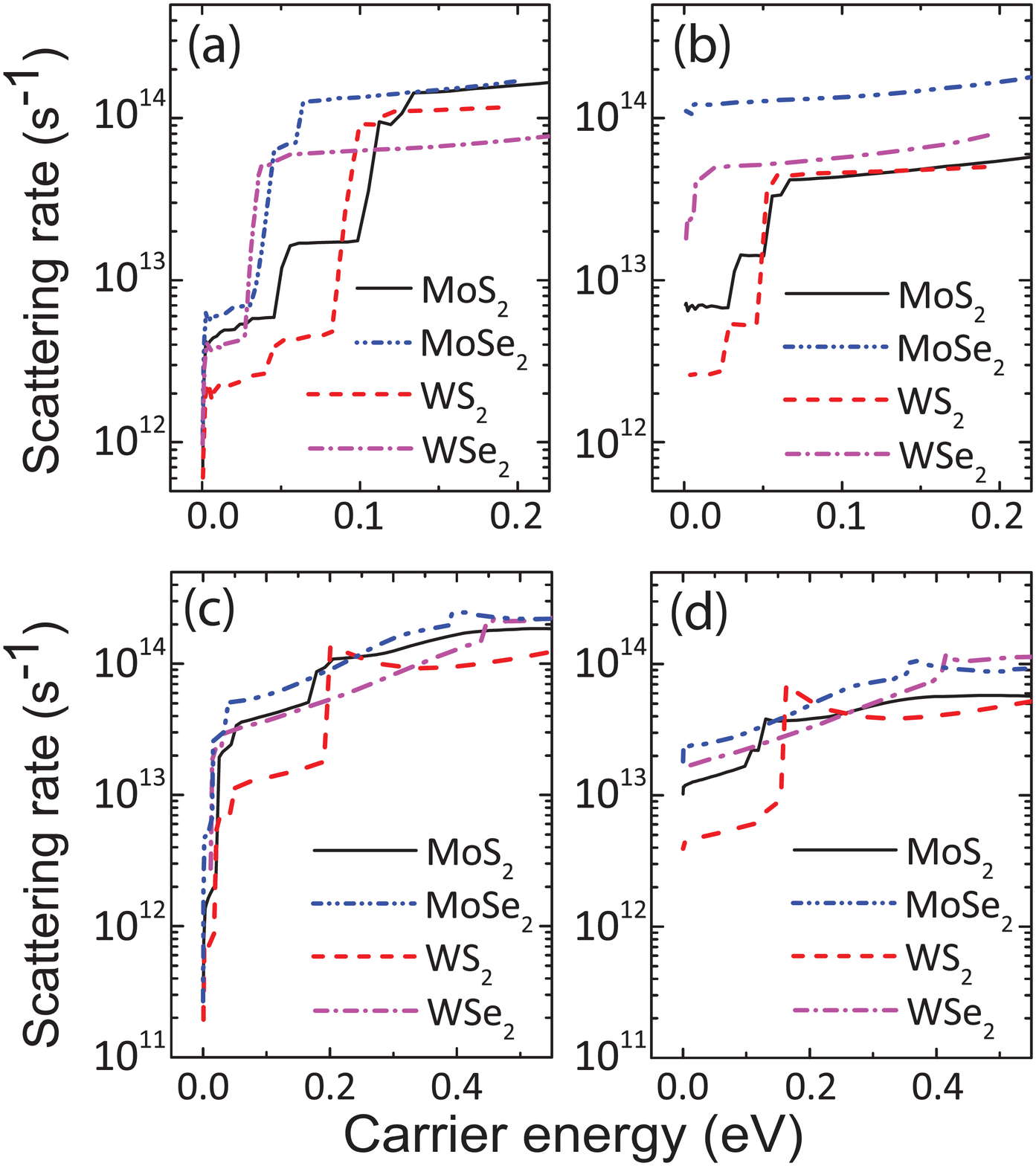} 
\caption{(Color online) Total intrinsic  scattering rates of (a,b) K-valley electrons and (c,d) K-valley/peak holes in four monolayer TMDs at room temperature. (a) and (c) are for phonon emission, while (b) and (d) correspond to phonon absorption.  The electron/hole wave vector $\mathbf{k}$ is assumed to be along the K-$\Gamma$ axis as the carrier energy increases. }
\end{figure}
\end{center}

\clearpage
\begin{center}
\begin{figure}
\includegraphics[width=8cm]{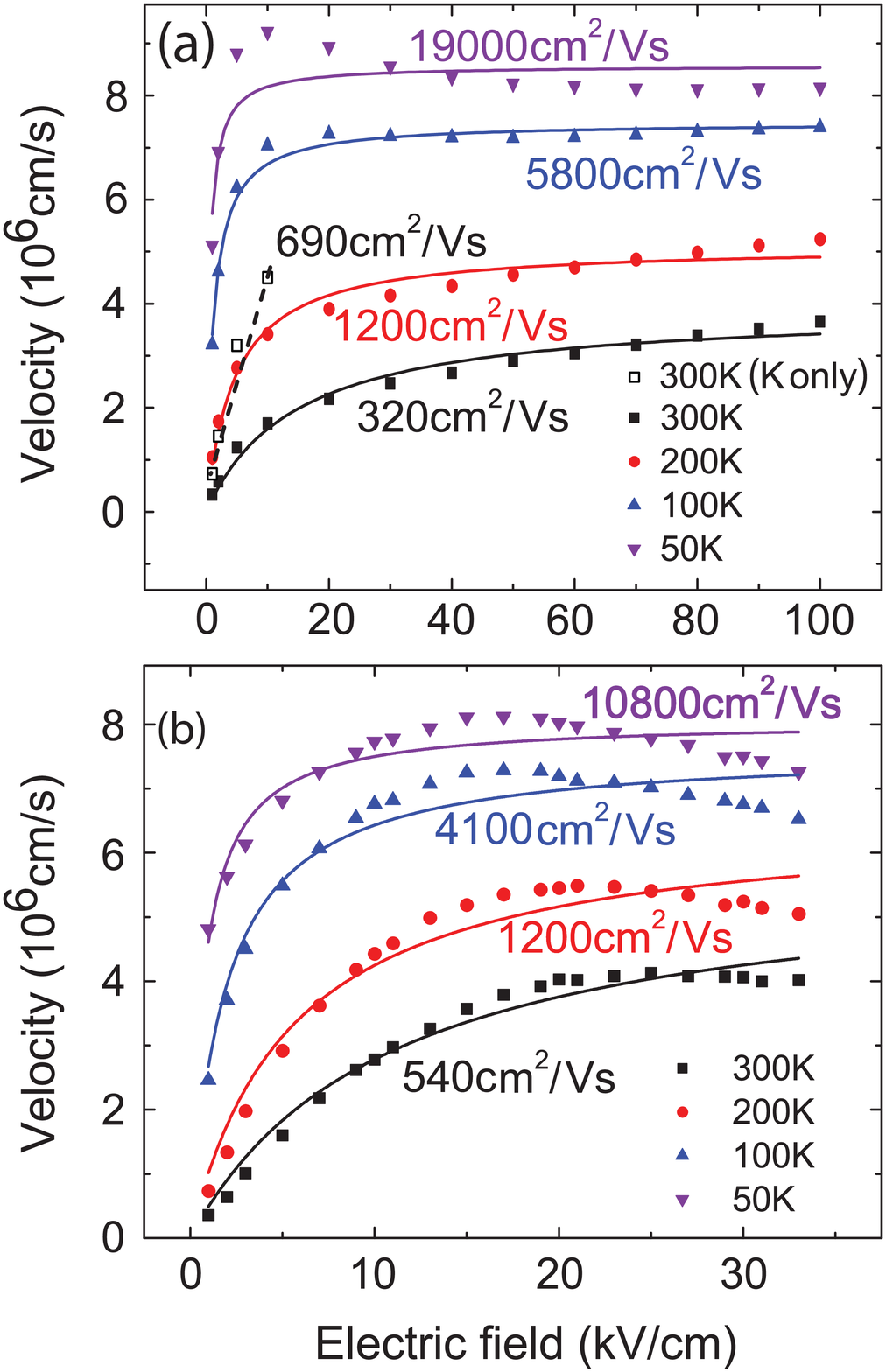} 
\caption{(Color online) (a) Electron and (b) hole drift velocity versus electric field in monolayer WS$_2$ obtained by a full-band Monte Carlo simulation at different temperatures. When electron transfer to the Q valleys is not considered, the mobility increases to approximately 690 cm$^2$/Vs at 300 K [represented by the dashed line in (a)].}
\end{figure}
\end{center}

\clearpage
\begin{center}
\begin{figure}
\includegraphics[width=15cm]{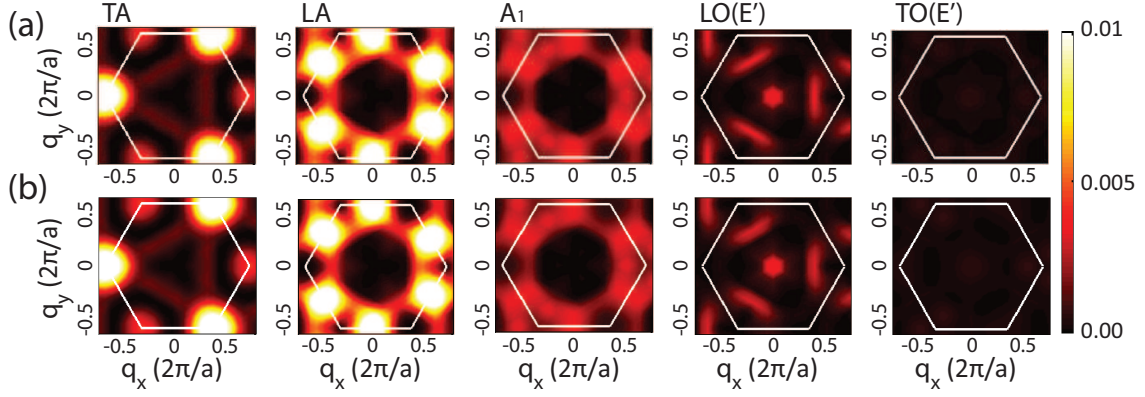} 
\caption{(Color online) Squared carrier-phonon interaction matrix elements $|g_{\textbf{q},\textbf{k}}^{v}|^2$ (in units of eV$^2$) in WS$_2$ as a function of phonon wave vector $\mathbf{q}$ for the initial electron state $\mathbf{k}$ at the valence-band maximum K point.  In the DFT/DFPT calculation, (a) full relativistic and (b) nonrelativistic psuedopotentials are used with and without the effect of spin-orbit coupling, respectively.  For (a), the plot shows the averaged $|g_{\textbf{q},\textbf{k}}^{v}|^2$ ($\propto 1/\tau$) between the two spin split bands present at the same $\mathbf{k}$ state.  Only the phonon branches with significant contribution are shown. }
\end{figure}
\end{center}

\end{document}